A break from the norm? Parametric representations of preference heterogeneity for discrete choice models in health.

John Buckell, Nuffield Department for Population Health and Nuffield Department for Primary Health Care Sciences, University of Oxford

Alice Wreford, School of Health Sciences, University of East Anglia

Matthew Quaife, Evidera

Thomas Hancock, Institute for Transport Studies and Choice Modelling Centre, University of Leeds and Nuffield Department for Primary Health Care Sciences, University of Oxford

Abstract

Background: Any sample of individuals has its own, unique distribution of preferences for choices that they make. Discrete choice models try to capture these distributions. Mixed logits are by far the most commonly used choice model in health. A raft of parametric model specifications for these models are available. We test a range of alternatives assumptions, and model averaging, to test if or how model outputs are impacted.

Design: Scoping review of current modelling practices. Seven alternative distributions, and model averaging over all distributional assumptions, were compared on four datasets: two were stated preference, one was revealed preference, and one was simulated. Analyses examined model fit, preference distributions, willingness-to-pay, and forecasting.

Results: Almost universally, using normal distributions is the standard practice in health. Alternative distributional assumptions outperformed standard practice. Preference distributions and the mean willingness-to-pay varied significantly across specifications, and were seldom comparable to those derived from normal distributions. Model averaging offered distributions allowed for greater flexibility, further gains in fit, reproduced underlying distributions in simulations, and mitigated against analyst bias arising from distribution selection. There was no evidence that distributional assumptions impacted predictions from models.

Limitations: Our focus was on mixed logit models since these models are the most common in health, though latent class models are also used.

Conclusions: The standard practice of using all normal distributions appears to be an inferior approach for capturing random preference heterogeneity.

Implications: Researchers should test alternative assumptions to normal distributions in their models.

Highlights

- Health modellers use normal mixing distributions for preference heterogeneity
- Alternative distributions offer more flexibility and improved model fit
- Model averaging  offers yet more flexibility and improved model fit

Distributions and willingness-to-pay differ substantially across alternative

Introduction

Discrete choice models are used widely across health economics to answer questions on health behaviours and clinical choices, to inform the development of Randomized Controlled Trials (RCTs), for policy choices and public opinion, and in economic evaluation. Hundreds are published each year (Soekhi et al., 2021). Policymakers such as the National Institute for Health and Care Excellence (2021), Canadian Agency for Drugs and Technologies in Health (2017), and the Food and Drug Administration (2016) are increasingly using evidence from these methods.

Broadly speaking, there are three acknowledged types of heterogeneity in choice behaviour: *preference heterogeneity* (differences in preferences for attributes and alternatives), *scale heterogeneity* (differences in the randomness in choices), and *decision rule heterogeneity* (the decision rules/heuristics that individuals use when making choices). In health economics, attention has been given to scale heterogeneity (Wright et al., 2018; Buckell and Hess, 2019) and to decision rule heterogeneity (e.g. Erdem et al., 2014; Buckell et al., 2022; Meester et al., 2023), but the overwhelming attention has been in attempting to capture preference heterogeneity (Vass et al., 2022). Here, standard practices have emerged over the last two decades, which we bring into question.

Preference heterogeneity can either be classed as deterministic or random. Deterministic preference heterogeneity involves relating choice behaviour to observed individual characteristics. Random preference heterogeneity is that which is unobserved to the analyst, and can be modelled in a variety of ways, broadly divided into discrete and continuous mixtures (see Figure 1). In this paper, we focus on the specification of mixing distributions in continuous mixture models which are the most frequently implemented choice model in health economics. As per Figure 1, there are many other approaches that researchers could adopt, including allowing for correlation across mixing distributions (Revelt and Train, 1998), latent classes (Hess, 2014), mixed latent class models (Hensher and Greene, 2013), logit mixed logit models (Train, 2016), and nonparametric approaches (Vij and Kreuger, 2017). Researchers may combine these structures and/or average over them. In this paper we highlight some approaches and compare them the researcher norms in the field.

[Insert fig. 1 here]

An overview of how heterogeneity can be modelled within discrete choice models. We specifically focus on the distributional forms assumed within mixed logit models. NB - this list is not exhaustive: other options exist.

Examples of both approaches abound in health, though recent evidence shows that analysts in health are using continuous mixtures more than any other model (Vass et al., 2022). Of these continuous mixtures, recent evidence indicates that studies in health are overwhelmingly assuming normal distributions to model preference heterogeneity (Vass et al., 2022). We extended this analysis with a scoping review assessing current practices of modelling unobserved preference heterogeneity in health-based choice modelling, corroborating these findings.

In recent years, there have been major advances in the way that choice modellers have been able to capture random preference heterogeneity. These include transformations of basic distributions to impose constraints on preferences, such as loguniform distributions which impose a directionality of preference (Bhat, 1998; Hess et al., 2005); and more flexible distributions that allow for asymmetry such as asymmetric triangular (Dekker, 2016) or using transformations of normal as in Fosgerau and Mabit (2013). Further to this is model averaging (Hancock and Hess, 2021; Hancock et al., 2020) which can potentially improve on individual specification and, more importantly, mitigates the risk of analyst bias (for example observer bias where the choice of distribution might bias estimates (Porta, 2014)). It additionally can protect against overfitting, with Hancock et al., (2020) demonstrating that model averaging performs particularly well in forecasting, likely a result of at least one constituent model accurately predicting choices made by each individual in the holdout sample.

Two further issues that arise with sole reliance on normal distributions concern willingness-to-pay (WTP). First, as might be expected, if the normal distribution is not a reasonable depiction of the true preference heterogeneity, then bias can be introduced in the distribution of WTP (Tabasi et al., 2024). Importantly, even if the research interest is only in the mean, rather than the distribution, of WTP, this too could be biased. This is potentially highly problematic if results from these studies are used elsewhere (e.g. in cost-benefit analysis). Second, normal distributions can prevent moments of marginal WTP distributions (Daly et al., 2012), and alternative distributions or indeed specifications are used elsewhere for estimating this instead (Train and Weeks, 2005; Crastes dit Sourd, 2024).

We use four datasets in analyses to examine alternative specifications of continuous mixing distributions. Two are stated preferences: smoking choices in the US and HIV prevention choices in South Africa. These were chosen due to known preference heterogeneity. For example, it is well known that preference for menthol flavours vary amongst smoker groups, implying that the preference distribution is likely asymmetric (Buckell et al., 2019). A further dataset is a revealed preference analysis of smoking and vaping in the US. This was chosen to demonstrate the applicability of these methods in modelling real-world choices, and to extend the modelling framework from the multinomial logit to multivariate logit models with correlated errors. A simulated dataset of drug choices completes the set. An advantage is that this data can be shared, along with the code script, for replicability and use by researchers in the field.

Using these datasets, we compare standard practice with alternative representations of continuous mixing distributions. More specifically, we examine model fit, choice probabilities, WTP means and preference distributions. We reconcile our scoping review with our empirical exercises to ask whether current practices can be improved; and the extent to which this is important for empirical measures derived from choice models in health. By comparing standard practices with more advanced approaches, we provide evidence on the robustness of modelling results in health-based choice modelling.

The remainder of this paper is structured as follows. In the next section, we present our scoping review. We then present the datasets used in exercises. Next, we set out our choice modelling methods and strategy for making model comparisons. We then move to the modelling results. Finally, we discuss our findings and present our conclusions.

Methods

*Sampling and datasets*

1a. Smokers' stated preference tobacco product choice data, USA

Data were taken from an online DCE on 2,031 US adult smokers conducted in 2017 (1,531 current smokers; and 500 self-reported recent quitters) (Buckell et al., 2019). Sampling was based on quotas derived from the Behavioural Risk Factor Surveillance System (BRFSS) data in 2013/14, comprising gender, age, education and region to make the sample representative. The sample size is well in excess of minimum sample size calculations (de Bekker-Grob et al., 2015). A series of exercises were conducted to promote the quality of the data (e.g. attention checks in the survey, minimum time threshold, and removing duplicates individuals).

The DCE was based on a review of the literature and a pilot study. The literature review comprised prior DCEs in tobacco (e.g. a systematic review by Regmi et al., 2018); market data on tobacco product prices (Cuomo et al., 2015); and scientific literature on the harms of tobacco products (Jha et al., 2013; McNeill et al., 2015). In the study, individuals chose between cigarettes, e-cigarettes and opt-outs. Respondents were presented with 2 of each product and made two choices in each choice task. Attributes (levels) were price ($4.99, $7.99, $10.99, $13.99), flavours (tobacco, menthol, fruit, sweet), level of nicotine (none, low, medium, high) and health harm expressed in life years lost to the average smoker (2 years, 5 years, 10 years, unknown). Some levels were omitted to make choices realistic (e.g. fruit/sweet cigarettes are not on the market in the US). This design is based on reality (non-menthol tobacco flavours being banned), a review of the literature and a pilot study; full details can be found in Buckell et al.. 2019.

A Bayesian D-optimal design was used (Hensher et al., 2015). Priors were obtained from a multinomial logistic (MNL) model in analysis of pilot study data on 87 respondents. 3 blocks of 12 had individuals randomised to them. Each individual answered 12 choice sets, balancing concerns of learning and respondent fatigue (Hess et al., 2012). A practice choice scenario was given to all respondents to ensure that they understood how the choice scenarios worked.

1b. Smokers' revealed preference tobacco product choice data, USA

Smoking behaviour data was collected from the 2,031 sampled individuals. Each was asked on their use of cigarette and e-cigarettes. 1,038 reported cigarette use only, 148 reported e-cigarette use only, and 619 reported the use of both products; 226 reported that they had recently quit. Data on individuals' characteristics was collected alongside.

2. HIV prevention stated preference product choices among a general population sample, South Africa

In 2015, 367 HIV negative women (199 aged 16-17 and 168 aged 18-49) were interviewed in a randomised face-to-face household survey conducted in a peri-urban township on the outskirts of Johannesburg, South Africa (Quaife et al., 2017).

The DCE was developed through an analysis of a previous DCE and focus groups discussions carried out in previous research (Terris-Prestholt et al., 2014), specifically identifying important characteristics of prevention products and exploring optimal ways to present these in a clear and relatable manner to participants. This was supplemented by a scoping literature review to identify new products and additional attributes which could be important to respondents, which was added to and refined through piloting. Three alternatives of new products were shown in each task using an unlabelled design where each alternative represents a generic product within which all characteristics can change as prescribed by the statistical design. In this experiment, respondents chose between three unlabelled alternatives of new HIV prevention products and an opt-out. Products were described by product type (oral pill, injectable, reusable diaphragm, vaginal gel, and vaginal ring), HIV prevention efficacy (55%, 75%, 95%), contraceptive ability (yes, no), STI protection (yes, no), frequency of use (coitally, daily, weekly, monthly, every three months, every six months, annually), and side-effects (nausea, stomach cramps, dizziness, none). A Bayesian D-optimal design was generated using priors estimated on a MNL model in a pilot using a sequential orthogonal design.

3. Simulated drug choices data

Simulated drug choice data were generated for 1,000 individuals, who each 'completed' 10 choice tasks. In each task, two branded alternatives and two unbranded alternatives were presented, described by country of production, characteristics of the drug (standard, fast acting or double strength), risk of side effects and price. The attribute levels were based on the example choice dataset given on the Apollo choice modelling website. The choices were generated using random draws U[0,1], with the probability of choosing each alternative generated using a MNL model. The utility for each alternative was defined by specifying a taste for each attribute for each individual (drawn from distributions), where different underlying distributions were used for different attributes.

*Choice modelling*

Random utility maximisation (RUM) models has been used almost exclusively for choice models in health (Buckell et al., 2022; Meester et al., 2023). In this formulation, the individual reconciles their product/attribute preferences for each of the available alternatives and chooses that which maximises their utility. Respondents' utility is typically specified by the modeller as a linearly-additive function of attribute/alternative preferences and the alternative-attribute combinations available, with an error term to capture noise. For each alternative, the individual is assumed to choose the option that delivers the highest utility.

$$U_{nti} = V_{nti} + \varepsilon_{nti} \qquad (1)$$

Where $U_{nti}$ is the utility for decision maker *n* for alternative *i* in choice task *t*, comprising deterministic (observed) and random (unobserved) utility. $V_{nti}$ is the deterministic component of utility; and $\varepsilon_{nti}$ is the random component of utility (Train, 2009), capturing the fact that the analyst does not observe all factors that may lead to a decision, where factors may vary across individuals, alternatives or specific choice tasks.

For the revealed preference data, logit models are specified on both outcomes.

$$U_{cig,n} = V_{cig,n} + \rho_n + \varepsilon_{cig,n} \quad (2)$$

and

$$U_{ecig,n} = V_{ecig,n} + \rho_n + \varepsilon_{ecig,n} \quad (3)$$

Where $V_{cig,n}$ is the deterministic component of utility for cigarette use and $V_{ecig,n}$ is the deterministic component of utility for e-cigarette use. $\rho_n$ is an individual-specific error component capturing correlation across the errors for both outcomes.

*Choice probabilities*

For stated preference and simulated data, the random utility maximisation (RUM) model is operationalised by assuming a type-I extreme value error distribution on $\varepsilon_{nti}$ for each alternative and estimating choice probabilities for each alternative, resulting in the multinomial logit (MNL) model (Maddala, 1983).

$$P_{nti} = \frac{\exp{(V_{nti})}}{\sum_{j=1\ldots J} \exp{(V_{ntj})}} \quad (9)$$

Where $P_{nti}$ is the RUM probability of respondent *n* choosing alternative *i* from set *J* in a choice task. The probability of individual *n* making a sequence of choices, each *t*, over the total, $T_n$, choice tasks that they face is then:

$$P_n = \prod_{t=1}^{T_n} P_{ntj^*} \quad (10)$$

Where *j*\* is the alternative selected in each given scenario.

For the revealed preference data, assuming a type-I extreme value error distribution on $\varepsilon_{n,cig}$ and $\varepsilon_{n,cig}$ gives,

$$P_{n,cig} = \frac{\exp(V_{n,cig}+\rho_n)}{1+\exp(V_{n,cig}+\rho_n)} \quad (11)$$

And

$$P_{n,ecig} = \frac{\exp(V_{n,ecig}+\rho_n)}{1+\exp(V_{n,ecig}+\rho_n)} \quad (12)$$

Where $P_{n,cig}$ is the probability of reported cigarette use, and $P_{n,ecig}$ is the probability of reported e-cigarette use

Thus, we have:

$$P_n = (P_{n,cig})^{c_n} \cdot (1 - P_{n,cig})^{(1-c_n)} \cdot (P_{n,ecig})^{e_n} \cdot (P_{n,ecig})^{(1-e_n)} \quad (13)$$

Where $c_n$ and $e_n$ are dummy variables that take a value of 1 if individual *n* is a smoker or e-cigarette smoker, respectively.

Model log-likelihoods are given by,

$$LL = \sum_{n=1}^{N} \ln P_n \quad (14)$$

*Model averaging*

The approach here follows the sequential latent class approach set out in Hancock et al. (2020) and Hancock et al. (2021). This formulation adapts a standard latent class framework such that the classes are the set of models over which model averaging occurs, and the class membership probability is the weight applied to each model. The constituent models – each treated as a class in the LC model – are estimated separately in a prior estimation stage. They are then entered into the latent class framework with the parameter estimates for the models held constant whilst the class shares are estimated. Specifically, supposing there are *K* models, each has a set of estimated parameters that we denote $\widehat{\Omega}_k$. Then, the only parameters to be estimated, $\theta_k$, are those that feed the class membership probabilities, $\pi_k$, which is the averaging. That is, the estimated class membership probability parameters govern the weights for each constituent model and optimise model fit. Thus, the model averaging will give more weight to constituent models with superior fit.

$$\pi_k = \frac{\exp(\theta_k)}{\sum_{k=1}^{K} \exp(\theta_k)} \quad (15)$$

Where the logistic form ensures $\sum_{k=1}^{K} \pi_k = 1$ and $0 \leq \pi_k \leq 1 \; \forall k$.

Combining the components leads to the log-likelihood for sequential latent class model averaging,

$$LL_{MA}(\pi_k, \widehat{\Omega}_k) = \prod_{n=1}^{N} \ln\left(\sum_{k=1}^{K} \pi_k \cdot P_n(\widehat{\Omega}_k)\right) \quad (16)$$

Whilst it is possible to average over any set of models, we restrict our modelling in this setting to three groups of models, and our presentation to one group containing all constituent models as that yielded the largest gain in fit and flexibility. The groups of models were: the base models MA(S,U,T); base + extended models (S,U,T,LN,LU); base + extended + flexible models MA(S,U,T,LN,LU,FM2,FM3).

*Application of choice models to datasets*

The deterministic component(s) of utility is then defined for each dataset. For tobacco stated preference data,

$$V_{nti} = ASC_{cig,n}.Cig_{nti} + ASC_{ecig,n}.Ecig_{nti} + \beta_{p,n}.Price_{nti} + \beta_{N,n}.Nicotine_{nti} + \beta_{f,n}.Flavor_{nti} + \beta_{h,n}.Heath\ Harm_{nti} \quad (4)$$

Where $V_{nti}$ includes alternative-specific constant terms (ASCs) that are added if alternative *i* is a cigarette ($ASC_{cig,n}$) or an e-cigarettes ($ASC_{ecig,n}$), where the opt-out is the reference product. We then have attributes of price, nicotine, flavours, and health harm expressed in the number of life years lost; and corresponding preferences ($\beta$) which vary over individuals, *n*, according to distributions (cf. Table1). This model is also estimated in the willingness-to-pay space (Train and Weeks, 2005), which avoids undefined moments in the WTP distribution (Daly et al., 2012).

$$V_{nti} = \beta_{p,n}.(Price_{nti} + ASC_{cig,n}.Cig_{nti} + ASC_{ecig,n}.Ecig_{nti} + \beta_{N,n}.Nicotine_{nti} + \beta_{f,n}.Flavor_{nti} + \beta_{h,n}.Heath\ Harm_{nti}) \quad (4a)$$

For HIV prevention,

$$V_{nti} = ASC_{optout,n}.Optout_{nti} + \beta_{p,n}.Product\ type_{nti} + \beta_{pro,n}.Protection\ from\ diseases_{nti} + \beta_{pre,n}.Prevention\ pregnancy_{nti} + \beta_{freq,n}.Frequency\ of\ use_{nti} + \beta_{se,n}.Side\ effects_{nti} \quad (5)$$

Where $V_{nti}$ includes an alternative-specific constant for the opt-out; and attributes of product type, protection from diseases, pregnancy prevention, frequency of use, and side effects with corresponding preferences ($\beta$) which vary over individuals, *n*, according to distributions (cf. Table1).

For simulated drug choices,

$$V_{nti} = \beta_{b,n}.Branded_{nti} + \beta_{c,n}.Country_{nti} + \beta_{ch,n}.Characteristic_{nti} + \beta_{se,n}.Side\ effects_{nti} + \beta_{p,n}.Price_{nti} \quad (6)$$

Where $V_{nti}$ includes attributes of branded, country of origin, drug characteristic (e.g. fast acting), side effects, and price with corresponding preferences ($\beta$) which vary over individuals, *n*, according to distributions (cf. Table1).

For tobacco revealed preference,

$$V_{cig,n} = ASC_{cig,n} + \gamma_{cig,n}.z_n \quad (7)$$

And

$$V_{ecig,n} = ASC_{ecig,n} + \gamma_{ecig,n}.z_n \quad (8)$$

Where $V_n$ includes an alternative-specific constant for cigarettes/e-cigarettes, and individual characteristics, $z_n$ with corresponding parameters ($\gamma$) which vary over individuals, *n*, according to distributions (cf. Table1).

In all specifications, attribute levels are dummy-coded (which is equivalent to effects-coding, as used widely in health; see Daly et al., 2016) except for price which is treated continuously.

The utility functions above can then be extended to accommodate deterministic and random heterogeneity. Given our interest in the latter, our discussion focusses on this. Random heterogeneity can take two main forms, namely discrete mixtures (i.e. latent classes; see Hess (2014)) and continuous mixtures (mixed logit models; see Train (2009)). Latent classes generally alter all of the coefficients (though need not), allowing for a number of classes each with its own fixed set of parameter estimates. Mixed logit models specify the preference for some or often all attributes as a distribution, which can take a wide range of forms, a point that we next consider. Mixed latent class models combine both structures, where a set of classes are estimated and distributions of preferences are specified within each class.

*Distributional assumptions in mixed logit models*

Given the freedom with which researchers can specify mixed logit models, there are an almost infinite number of possible model specifications. Thus, a comparison across all sets of possibilities is unwieldy. To make a reasonable set of models, and then comparisons across models, three classes of models are defined in the current work. We begin with what we refer to as a "basic" set which includes the standard (N) approach in current practice in health; that is, setting all mixing distributions to be normal (see Table 1). We next use two further "basic" distributional assumptions that are used in the choice modelling literature (predominantly outside of health research), namely uniform (U) and triangular (T). We next define a group of "extended" basic distributions, where we take transformations of two basic forms to allow for asymmetry in the distributions of preferences. Here, we have lognormal (LN) and log uniform (LU) distributions. LN and LU are desirable specifications for when strictly positive or negative preferences are to be imposed (though in this case we use only negative transformations). Finally, we use a group of "flexible" distributions which are the asymmetric triangular (AT) and the polynomial expansions set out in Fosgerau and Mabit (2013) to allow for further flexibility. Specifically, we use second order (FM2) and third order (FM3) specifications. The mathematical form and implementation of these distributional assumptions are set out in Table 1 below.

[Insert table 1 here]

*Model Estimation*

All models are estimated using the Apollo package in R (Hess and Palma, 2019). Models used 500 MLHS draws, except for the tobacco RP dataset which used 100. This was due to the difficulty in estimating these models, where 100 draws made for more stability in estimation (we recognise that this may be too few – see limitations). For each model, comparisons are made across the model fit (log-likelihood) and the estimated choice probabilities for each model. Choice probabilities are computed using sample enumeration (Train, 2009). Unconditionals are constructed from the fitted model post-estimation and used for analyses of preference/WTP distributions. For model averaging, unconditionals are sampled from each constituent model according to the weights (derived from the estimates of $\theta_k$). The estimation of model averaging is thus a simple process (it only requires model outputs $P_n$ from each model) and the analyst need not know the underlying model that generated $P_n$ (Hancock et al., 2020).

Codes and simulated data are available on GitHub at: https://github.com/johnbuckell/Modelling-random-preference-heterogeneity-in-health-choices

Results

Scoping review

Table A1 shows the results for types of distributions used in mixed logit models retrieved in our search. In 2017, 98% (226/230) of all distributions were normal with 2% (4/230) being lognormal. In 2022, 99% (736/746) of all distributions were normal with 1% (10/746) being lognormal. Based on this, we define a "standard practice" in health to be using normal distributions for all parameters. Notably, almost half of papers did not report the distributional assumptions used in their model. This was part of a worrying theme of not reporting essential information on choice models, with many papers also omitting basic outputs/inputs such as model fit, types of draws used, software used or whether any model selection process had been undertaken (see Appendix 1 for full information; and Table A1 for full results).

Choice modelling

Figure 2 (and Table A2) shows the results from the set of 8 models and model averaging for each of the four datasets. For AIC, lower values denote a better fit, and so values further to the left are superior. In terms of model fit, there are substantial differences across models in 3 of 4 datasets (tobacco SP, HIV prevention, and simulated drug choices), and minimal differences in one dataset (tobacco RP). Further, the asymmetric triangular distribution was not estimable on this data; the model collapsed to the triangular distribution. For the tobacco SP data, we extended analyses to the willingness-to-pay space; Appendix 3 and Table A3 show results for these estimates.

The first analysis is the comparison of model fit (here, the AIC) of standard practice, all normal (S), with alternative distributional assumptions. Standard practice is outperformed in two of four cases by alternative base models (simulated drug choices and tobacco RP data), in three of four cases by the extended models (tobacco SP, HIV prevention, and simulated drug choices), in three of four cases for extended models (tobacco SP, HIV prevention, and simulated drug), and in one of four cases by the flexible models and model averaging (tobacco SP). In all cases, moving away from standard practice resulted in a better model fit.

An additional analysis reran the standard practice model for tobacco SP omitting each attribute singly as yardsticks against which to pitch gains in fit of alternative distributions. We found losses in LL of: 363 units (health harm), 2,320 (flavour), 279 (nicotine), and 1,768 (price). These compared to a difference of 685 units between standard practice and the best fitting model, MA3. Hence, improving choice of distributional assumption, even in our rather limited form (by assuming all attributes follow the same distribution), improves fit by a comparable amount as a low explanatory attribute.

The second analysis is on the models' predicted choice probabilities. Table 2 compares the predicted choice probabilities for each model and for model averaging on each of the four datasets. There does not appear to be much, if any, impact of either distributional assumptions or model averaging on forecasts from models. Overall, it does not appear that distributional assumptions impact on models' predicted choice probabilities.

The third analysis is of WTP, shown in Fig. 3, with menthol flavour (reference: tobacco flavour) and e-cigarette (reference: the opt-out in the experiment) preferences taken as examples. These are taken from the WTP space model and are directly estimated without the need for post-estimation calculation of point estimates and standard errors. There is considerable heterogeneity across the mean WTP for both parameters. For menthol, the WTP for the normal distribution is $-5.90 (95%CI: $-5.21 to $-6.58), which also happens to be the highest. The minimum, from the FM3 model, $-12.78 (95%CI: $-14.38 to $-11.17), is around double that from the normal distribution and statistically significantly different. An FM4 model ($-16.85) was tested and omitted as it had worse BIC, but suggests that the estimate for FM3 is not an outlier, perhaps better capturing the tail (implying many individuals have a strong negative preference for menthol, see Figure 3). The model averaging WTP, $-9.35 (95%CI: $-10.51 to $-8.19), is also considerably lower than that of the normal distribution and statistically significantly different. For e-cigarettes, the WTP for the normal distribution, $7.83 (95%CI: $7.07 to $8.59), is neither the highest nor lowest. The WTP from the FM3 is lowest, $6.19 (95%CI: $5.19 to $7.21), and WTP from the loguniform model is highest, $13.02 (95%CI: $12.32 to $13.73); both are statistically significantly different from the WTP from the normal distribution. The model averaging WTP, $7.13 (95%CI: $6.27 to $7.99), is also lower than that of the normal distribution though not statistically significantly different. As expected, the choice of distributional assumption has important ramifications for the estimates of WTP. WTP was different and for menthol, statistically significantly different for the preferred model (according to AIC) relative to standard practice.

A fourth analysis of preference distributions is presented in Fig. 4. This shows the probability densities for e-cigarette (reference: the opt-out in the experiment) and menthol (reference: tobacco) preferences. The shapes of distributions for the base models resemble the impositions made on them. The shapes of the preference distributions in the extended models differ, and embody the unidirectionality imposed by their specifications. In the flexible models, the shapes of the distributions are substantially different and allow for asymmetry in the preference distributions, and multimodality in the FM models. For model averaging, the shape of the distribution reflects the model specification in that they are weighted averages of the probability densities of, i.e. distributions imposed by, the constituent models. The shapes of the distributions are similar for some models (e.g. normal versus triangular), and very different for others (e.g. normal versus FM3). For preferred models, the shapes of distributions are very different to those recovered from applying standard practice.

Finally, the recoverability of underlying preference distributions in the simulated dataset are illustrated in Figure 5. This figure shows (a) the underlying distribution, (b) the fitted normal distribution, and (c) the distribution of the best-fitting model (MA). MA, given its flexibility, captures these distributions reasonably well. In particular, it provides an accurate account of the preferences for 'branded' which is multimodal. This was not the case for the normal distributions. Normal distributions seemed to only fit well when the underlying preferences were themselves normal. Unsurprisingly, normal distributions do not fit non-normal shapes very well. It performed poorly elsewhere. This demonstrates the effectiveness of model averaging over more complex distributions in capturing the underlying distributions.

[Insert fig. 2 here]

Fig 2: AIC of models and model averages over four datasets. Model averages combine all of the models in each dataset, that is MA3(S,U,T,LN,LU,AT,FM2,FM3). NB – a conservative approach to AIC for model averaging of counting all parameters from all constituent models, as opposed to only the parameters estimated in the second stage of estimation.

[Insert table 2 here]

[Insert fig. 3 here]

Fig 3: Estimates of WTP for e-cigarettes and menthols flavour on the tobacco SP data. Estimates derived from the WTP space model (eqn. 4(a)).

[Insert fig. 4 here]

Figure 4: Illustration of preference distributions from tobacco SP models for e-cigarettes and for menthol, across different distributional assumptions and model averages.

[Insert fig. 5 here]

Figure 5: Distributions of preferences for different attributes in the simulated drug choice data. The underlying distribution is in green and varies across attributes. The distribution from model averaging is given in red and the normal distribution is in blue. This figure uses the MA3(S,U,T,LN,LU,AT,FM2,FM3) model which draws from the full set of models.

Discussion

In this paper, we considered random preference heterogeneity in discrete choice models in health-based choice modelling. A scoping review of the literature established current practices in modelling (and reporting of modelling). There is no defined reporting standard for choice modelling. The most commonly used model in health is the mixed logit. Standard practice in health is to use normal mixing distributions for all parameters in models. We show that there are better alternatives that consistently fit the data better and have significantly different model outputs, implying that standard practice may give biased outputs.

With four datasets, 8 specifications of mixing distributions were compared, including standard practice, in multinomial logit and logit-based choice models. These ranged from simple assumptions to flexible approaches (the latter introduced here in health). We also used model averaging as a simple method to reduce analyst bias.

Alternative distributional assumptions offered some gains in model fit to standard practice in all four settings. Flexible approaches offered the largest gains in fit among individual models; model averaging improved model fit further in one of four cases.

Alternative distributional assumptions did not impact on in-sample predicted choice probabilities across the datasets studied. Model averaging likewise did not impact on choice probabilities (likely because it is drawing from these models). Note, however, that model averaging has been shown to improve out-of-sample forecasting in all datasets tested (Hancock et al., 2020; Chen and Liu, 2023), thus we opt not to repeat this test here.

Alternative distributional assumptions yielded a wide range of WTP estimates, many of which were statistically significantly different to those derived using normal distributions. (NB – the denominator in WTP is constant, implying that beta estimates vary also.)

Alternative distributional assumptions yielded a wide range of preference distributions, which allowed for both asymmetry and multimodality, which the normal distribution does not; they are further able to avoid the fact that normal distributions have long tails, which implies extreme preferences (an assumption which may not be warranted). Model averaging demonstrated good recovery of the underlying distributions in the simulated data, which was not the case for normal distributions.

Aside from predicted choice probabilities, these results raise serious concerns for standard practice in health-based choice modelling. Normal distributions were inferior specifications in terms of model fit in all cases. Not only did the preference distributions of alternative assumptions depart markedly from normal distributions, our simulations indicated that alternatives were far better able to recover the true underlying distributions in the data. The differences observed in WTP in alternative distributions suggests that those derived from models using normal distributions are likely to be biased to varying degrees. This brings into question the robustness of findings of standard practice.

Our scoping review revealed dominant use of pre-packaged software among those reported. This may be limiting if routines are not available for the full range of functional forms; and default settings may inadvertently dictate the choice of distribution. Free software, with code, is now available for researchers to use the alternatives studies here, and further specifications that are not (e.g. higher order polynomials).

Strengths of our study include a scoping review to document current practices and inform standard practice. We used four datasets comprising stated preferences, revealed preferences, and simulated data. We used the common multinomial logit model for health choices, but also extended our methods to include logistic regression with correlated errors. We introduced new flexible mixing distributions to health, as well as model averaging, which we have shown to offer substantial gains. The use of simulated data means that not only are we able to share code for estimating the models, but researchers can download the data and replicate these results.

Our study is subject to a set of limitations. First, we did not include models with latent classes in our modelling exercises. This is partly due to the primacy of mixed logit modelling in the literature, and partly due to keep the research questions focussed. As per our introduction, there are many other model structures that analysts could use: including allowing correlation across mixing distributions (Revelt and Train, 1998), latent classes (Hess, 2014), mixed latent class models (Hensher and Greene, 2013), logit mixed logit models (Train, 2016), and nonparametric approaches (Vij and Kreuger, 2017). Some of these, for example correlated mixing distributions and mixed latent classes, are easily implemented in software packages. We also did not use deterministic heterogeneity in our analyses, which is standard practice in choice modelling. This is in part to keep the exposition simple and the number of estimated models manageable, and in part reflecting current practices in health, which are infrequently to use deterministic heterogeneity. We further recognise that model averaging will require additional correction for standard errors given that there are two stages in estimation. We leave this issue for future research. We further note that that model averaging is limited to the space of its constituent models. We were limited to using so few as 100 or 500 draws in estimation. This was a consequence of limited computing power and in the case of the tobacco RP data, stability in estimates. It is well known that more draws yield more reliable results (Chiou and Walker, 2007; Czajkowski and Budzinski, 2019). We use only parametric approaches (and semi-nonparametric in the case of FM2 and FM3). There are nonparametric approaches that are also available and should be investigated (c.f. Tabasi et al., 2024). We were unable to estimate the AT model on the revealed preference data, which has a fairly large sample size of over 2,000 observations. Other revealed preference datasets with fewer observations may find difficulties in estimating the more complex specifications. Model averaging did not aid this issue in this setting. Finally, specifications in which all parameters had the same distributions applied to them were considered; distributions can of course be specified on a parameter-by-parameter basis, implying that further gains in fit may be possible.

We refrain from providing specific guidance in this paper for several reasons. Firstly, it is not quite clear what guidance could be given at all. For example, consider the case of advice on when to use log-transformed distributions, for example. A lot of choice modellers would use these for a cost attribute based on economic theory. We could extend this to health for, say, health harms (since it could be argued that preferences for these should be strictly negative). Yet, two issues arise. First, that some individuals may prefer more harmful products. For example, some hardened smokers may consider reduced harm products to be a sign of weakness (as per Thirlway et al's (2016) narratives) and therefore less appealing. Second, we may wish to first use a distribution that allows preferences to cross zero. Even contrary to expectations it still may be that a reasonable proportion of preferences cross zero, and that may be a signal of data quality issues (e.g. omitted variable bias). Secondly, we are concerned that guidance and suggestions are treated as direct instructions and may become norms. This may not be good for future research and in essence is the practice that we are trying to challenge here (i.e. using "all normals" is the current default practice). Third, that our results are varied and hence there is little opportunity to give guidance beyond "be sure to test the shape of the distribution that you use", which we hope is self-evident from our findings. In some cases, it should be noted that the distributions tested here are not helpful, as they may not have the required flexibility to accurately capture preferences (e.g. if we expect both positive and negative preferences for a given attribute, lognormals are clearly inadequate). Conversely distributions may have too much flexibility (e.g. infinite support) may be unrealistic. Case-specific domain knowledge is required.

If the goal of the research is forecasting behaviours, our results suggest that there is no imperative to move away from normal mixing distributions. However, this is seldom the case, and there are many settings where preferences, preference distributions, and WTP measures are required; for example using WTP for cost-benefit analyses or valuation of non-market goods. In these settings, the use of normal mixing distributions is highly questionable and likely yields inaccurate model outputs. Alternative approaches can outperform standard practice, better approximate underlying preference distributions, and yield more reliable measures of WTP.

Normal distributions are used for mixing distributions in choice models in almost all cases in health. This evidence suggests that alternative approaches can better capture preference distributions and produce more reliable estimates from choice models.

| Distribution name | Form | Implementation | Symmetrical | Unidirectional | Bounded support |
| --- | --- | --- | --- | --- | --- |
| Normal | $\tau_m \sim N(\mu_m, \sigma^2_{\tau,m})$ | $\beta_m = \mu_m + \sigma_m * d_{N,m}$ | Yes | No | No |
| Uniform | $\tau_m \sim U[a_m, a_m + b_m]$ | $\beta_m = a_m + b_m * d_{U,m}$ | Yes | No | Yes |
| Triangular | $\tau_m \sim T[a_m, a_m + b_m]$ | $\beta_m = a_m + b_m * (d_{U1,m} + d_{U2,m})$ | Yes | No | Yes |
| Lognormal | $\tau_m \sim LN(\mu_m, \sigma^2_{\tau,m})$ | $\beta_m = -e^{(\mu_m + \sigma_m * d_{N,m})}$ | No | Yes | No |
| Loguniform | $\tau_m \sim LU[a_m, a_m + b_m]$ | $\beta_m = -e^{(a_m + b_m * d_{U,m})}$ | No | Yes | Yes |
| Asymmetric Triangular | $\tau_m \sim AT[a_m, b_m, c_m]$ | $\beta_{m,lower} = a_m + (((a_m + b_m/2) + c_m) - a_m) * \sqrt{d_{U1,m}}$ <br> $\beta_{m,upper} = b_m - (b_m - ((a_m + b_m/2) + c_m)) * \sqrt{d_{U2,m}}$ | No | No | Yes |
| Fosgerau and Mabit | $\tau_m \sim \sum_{p=0}^{P} \alpha_{p,m} u^{p,m}$ | $\beta_m = \mu_m + \sum_{p=0}^{P} \sigma_{p,m} * d^{p,m}$ | No | No | No |

Table 1: Specifications of mixing distributions. $\mu_m$ is an estimated mean of a distribution; $\sigma_m$ is an estimated standard deviation of a distribution; $a_m$ is an estimated bound of a distribution; $b_m$ is an estimated range of a distribution; $c_m$ is an estimated offset; $d_{N,m}$ are draws from a standard (i.e. N(0,1)) normal distribution; $d_{U,m}$ are draws from a standard (i.e. U[0,1]) uniform distribution. In all cases, 500 Modified Latin Hypercube Sampling (MLHS, Hess et al., 2006) draws are taken. Estimation of the asymmetric triangular follows the procedure set out in Dekker (2016), where to reduce model runtime, $c_m$ is fixed to a value of 0 for an attribute if its inclusion does not significantly improve model fit. Note that for the applications in this work, tastes are assumed to vary across individuals, n. However, preferences may also vary across choice context, t, if we were to allow for inter and intra-respondent heterogeneity (Hess and Palma, 2023).

|  | | Tobacco SP | | | HIV prevention | | Drug choice simulated | | Tobacco RP | | | |
|---|---|---|---|---|---|---|---|---|---|---|---|---|
| Model | Description | cigarette | e-cigarette | opt-out | any product | opt-out | branded | unbranded | smoker | non-smoker | vaper | non-vaper |
| 1 | Normal | 0.50 | 0.38 | 0.13 | 0.68 | 0.32 | 0.46 | 0.54 | 0.82 | 0.18 | 0.38 | 0.62 |
| 2 | Uniform | 0.48 | 0.38 | 0.14 | 0.71 | 0.29 | 0.46 | 0.54 | 0.82 | 0.18 | 0.38 | 0.62 |
| 3 | Triangular | 0.50 | 0.37 | 0.13 | 0.68 | 0.32 | 0.44 | 0.56 | 0.82 | 0.18 | 0.38 | 0.62 |
| 4 | Lognormal | 0.48 | 0.37 | 0.15 | 0.66 | 0.34 | 0.45 | 0.55 | 0.82 | 0.18 | 0.38 | 0.62 |
| 5 | Log uniform | 0.50 | 0.37 | 0.13 | 0.66 | 0.34 | 0.46 | 0.54 | 0.82 | 0.18 | 0.38 | 0.62 |
| 6 | Asymmetric triangular | 0.51 | 0.36 | 0.13 | 0.67 | 0.33 | 0.47 | 0.53 | | | | |
| 7 | Fosgereau & Mabit ^2 | 0.50 | 0.37 | 0.13 | 0.64 | 0.36 | 0.46 | 0.54 | 0.82 | 0.18 | 0.38 | 0.62 |
| 8 | Fosgereau & Mabit ^3 | 0.49 | 0.38 | 0.13 | 0.63 | 0.37 | 0.46 | 0.54 | 0.82 | 0.18 | 0.38 | 0.62 |
| MA | Model averaging | 0.47 | 0.37 | 0.16 | 0.64 | 0.36 | 0.46 | 0.54 | 0.81 | 0.19 | 0.38 | 0.62 |

Table 2: Estimated choice probabilities for alternatives from 8 models and model averaging on four datasets.